# BMJ Open

# The socialisation of the adolescent who carries out team sports: a transversal study of centrality with a social network analysis

Pilar Marqués-Sánchez,[1] José Alberto Benítez-Andrades [ID],[2] María Dolores Calvo Sánchez,[3] Natalia Arias[1]

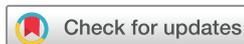

[1]Nursing and Physiotherapy, Universidad de León—Campus de Ponferrada, Ponferrada, Castilla y León, Spain
[2]Electric, Systems and Automatics Engineering, Universidad de Leon, Leon, Spain
[3]Administrative Law, Law Faculty, Universidad de Salamanca, Salamanca, Castilla y León, Spain

**Correspondence to**
Dr José Alberto Benítez-Andrades; jbena@unileon.es

## ABSTRACT

**Objectives** To analyse the physical activity carried out by the adolescents in the study, its relationship to being overweight (overweight+obese) and to analyse the structure of the social network of friendship established in adolescents doing group sports, using different parameters indicative of centrality.
**Setting** It was carried out in an educational environment, in 11 classrooms belonging to 5 Schools in Ponferrada (Spain).
**Participants** 235 adolescents were included in the study (49.4% female), who were classified as normal weight or overweight.
**Primary and secondary outcome measures** Physical Activity Questionnaire for Adolescents (PAQ-A) was used to study the level of physical activity. A social network analysis was carried out to analyse structural variables of centrality in different degrees of contact.
**Results** 30.2% of the participants in our study were overweight. Relative to female participants in this study, males obtained significantly higher scores in the PAQ-A (OR: 2.11; 95% CI: 1.04 to 4.25; p value: 0.036) and were more likely to participate in group sport (OR: 4.59; 95% CI: 2.28 to 9.22; p value: 0.000). We found no significant relationship between physical activity and the weight status in the total sample, but among female participants, those with overweight status had higher odds of reporting high levels of physical exercise (OR: 4.50; 95% CI: 1.21 to 16.74; p value: 0.025). In terms of centrality, differentiating by gender, women who participated in group sports were more likely to be classified as having low values of centrality, while the opposite effect occurred for men, more likely to be classified as having high values of centrality.
**Conclusions** Our findings, with limitations, underline the importance of two fundamental aspects to be taken into account in the design of future strategies: gender and the centrality within the social network depending on the intensity of contact they have with their peers.

## Strengths and limitations of this study

► Social network analysis is a tool that can help us understand the relational structure of the adolescent.
► Understanding the implications of the study of centrality within the social network of adolescents who engage in group sport is fundamental to the design of quality intervention strategies.
► The centrality of individuals from 11 different networks is analysed from a sociocentric perspective.
► Contact between individuals has been studied as a function of the time they spend together.
► The small sample size (n:235), the cross-sectional nature of the study and the network definition (classroom study) are the main limitations of this work.

20.6% and extreme or morbid obesity from 2.6% to 9.1% over the last 25 years.[2] In Spain, the figures from the latest National Health Survey (2017)[3] provide us with worrying data in the child and youth population group, as 28.7% of the population between the ages of 2 and 17 are affected by excessive weight (overweight+obesity), with the prevalence of being overweight or obese being highest among youth from households with relatively low socioeconomic status. Being overweight in adolescence is associated with low self-esteem and depression,[4] social isolation,[5] hypertension,[6] obesity hypoventilation syndrome,[7] musculoskeletal disorders,[8] mobility problems,[9] an increased risk of being overweight in adulthood[10–13] and onset of various forms of cancer.[14–17] Hence, it represents not only a health concern, but also a problem as regards the sustainability of health systems.

Poor diet and lack of physical exercise—the two main obesogenic factors—and factors that are associated with it, such as adolescent personal and contextual environment, can prompt an imbalance between intake and energy expenditure that leads to obesity.

## INTRODUCTION

The WHO considers obesity to be one of the most severe problems of the 21st century, classifying it as a 'global epidemic'.[1] In the USA, one of the most severely affected countries, adolescent obesity has risen from 10.5% to In Spain, for example, recent studies[18] have shown that 64.3% of children aged 6–17 have low or medium adherence to the Mediterranean diet and that 27.78% of young people aged 15–24 do not meet the minimum WHO recommendations for physical activity.[3] In relation to what was explained above, obesity can spread within an individual's social network[19] through acquiring ideas and modelling behaviours from our circle of friends and acquaintances,[20 21] or by being exposed to social influences that determine our way of being and acting in different areas of life.[22] In behaviours related to physical activity, several factors affect adolescent participation in sport,[23] but it is especially friends and peers that sometimes condition the amount of physical exercise performed in this age group.[24–29] In this context, Schofield et al[30] underlined the importance of friendship, noting that youth are most likely to experience social influence from the friends who they are most closely bound to—particularly when those friendships are reciprocal. Jago et al[31] conducted an exhaustive study of the relationship between physical activity in adolescence and sex and highlighted the importance of best friends for both sexes. Furthermore, Macdonald-Wallis et al[32] observed a difference in the influence exerted at a different level of friendship, that is, by friends of friends. However, as regards the study of changes in behaviour in the adolescent population and its mechanism of influence, the literature has widely described[19 33] that it can be brought about by different phenomena known as selection (homophilia) or contagion (influence). In fact, in the field of physical activity, both influence[34 35] and selection[32] have been demonstrated and that both modes can be present.[36]

The afore-mentioned have analysed contacts within a framework called network theory, using a methodology termed social network analysis (SNA), where the unit of analysis is the actor and his or her relationships. Born in the 1970s in the field of sociology and anthropology,[37] SNA has been used as a paradigm to analyse what relationships are like in a given social environment and how these relationships, as well as the positions that individuals occupy within networks, can be used to disseminate a change in health behaviour.[38] A comprehensive description of an individual's position in a network using specific variables describing network centrality (degree, indegree, outdegree, incloseness, outcloseness, betweenness and eigenvector) would be useful to obtain more information about of the social benefits of physical exercise. The common definition and interpretation of the afore-mentioned SNA measures used in this work is shown in table 1. The reason for using all possible measures of centrality and not opting for a single one was to enrich the existing literature and, using all measures, provide a

**Table 1** SNA measures used in this study

| SNA measure | Definition | Common interpretation | Impact on intervention |
|---|---|---|---|
| Indegree | Number of nominations an individual has received. | Leadership ability, social recognition or popularity. | Highly connected people are an important channel of information. The interventions carried out could affect a group of connected people but in return would not affect us less connected or important subgroups located on the periphery of the network |
| Outdegree | Number of nominations issued by an individual. | Ability to reach out to other adjacent individuals Sociability. Expansiveness | |
| Degree | Number of total connections an individual has. | Social activity | |
| Incloseness | Minimum distance from any node in the network to reach a given node. | Accessibility | People with a high degree of closeness could spread the information in the fastest way |
| Outcloseness | Minimum distance from a node to reach the rest of the nodes in the entire network. | Communication skills | |
| Betweenness | Number of times a student appears on the shortest path between any pair of peers. | Interpersonal influence | People with a high degree of intermediation control the flow of information. They are able to link groups, subgroups or individuals and are ideal when healthy behaviour is to be propagated. Conversely, they can be detrimental to negative health behaviours in the past. |
| Eigenvector | Eigenvector means that the centrality of each node is proportional to the sum of the centralities of the nodes it is adjacent to, in effect, when it comes to eigenvector centrality a node is a central as its network. | Degree of influence or prestige | Bonacich[91] suggests that power or recognition cannot be studied by traditional measures of centrality, and that this measure combines not only the degree of connection but also that of proximity. |

Source: Own elaboration based on[44 65 68 91–93].

more complete description of the centrality of individuals within the network in relation to the performance of physical exercise and in particular group sport. Furthermore, although SNA experts regularly use the degree as a measure of centrality in their studies because it is easy to understand and interpret,[39] there is no clear consensus as to the ideal measure of centrality as it depends on a particular context.[40] A number of studies into the application of the SNA paradigm to health interventions seek to identify the figure of the 'leader' or 'opinion leaders' who can act as facilitators of the spread of healthy or unhealthy behaviour within their network using different strategies.[41] For example, it has been shown that in the field of nutrition, according to indegree leaders, could spread unhealthy habits,[42] thus generating debate about the usefulness of this type of person as a promoter, as pointed out by Borgatti[43]. In fact, Borgatti[43] defends the use of the closeness centrality measure against that of indegree in the case of the dissemination of healthy habits, an idea which has been put into practice in the field of promoting physical activity.[44]

There is a lack of SNA studies that analyse the centrality (the position of individuals within the network) of individuals using adolescent's entire networks (sociocentric network perspective), especially in relation to obesity and physical exercise. The afore-mentioned centrality measures based on degree, betweenness and proximity are defined for 'complete' or 'sociocentric' network data that provide information on the relationships between all the nodes within a delimited social network. Since it is complicated to study the network from this study perspective when there are a large number of nodes, many researchers use the egocentric perspective in order to obtain information from a given node.[45] Therefore, analysing whole networks can extend the knowledge generated through egocentric analysis.

In this sense, the study of the structure of the social network of friends established by adolescents who participate in certain types of sports, such as group sports, is important in the field of obesity, since the adolescents concerned could obtain the necessary resources to be able to initiate physical activity or increase the physical activity carried out from their network of friends. This could be useful to help us understand their behaviour, and thus be able to design strategies based on it.

# THE STUDY
## Patient and public involvement
No patient involved.

## Aims
Since the social network of friends could modify certain habits, the purpose of this study is, on the one hand, to describe the physical activity carried out by the adolescents in the study, its relationship to being overweight (overweight+obese) and, on the other hand, to analyse the structure of the social network of friends established in adolescents doing group sports, by means of different parameters indicative of centrality.

## Design
The present study was conducted in the field of education. The students surveyed were in the third and fourth year of compulsory secondary education. A cross-sectional study involving 235 pupils aged 14–18 and belonging to 11 classroom networks was carried out in Ponferrada (Spain).

## Sample/Participants
The population invited to participate in this study was 776 students from 30 different classrooms (5 schools in total). Students were recruited through the teaching staff, aiming to achieve a participation rate of between 40% and 50% of class members. An initial sample of 276 students from 11 different classes was obtained. A participation rate per classroom ranging from 47.36% to 86.36% was achieved. The inclusion criterion applied to obtain our final sample consisted of individuals classified as 'normal weight', 'overweight' or 'obese' according to WHO references,[46] while those classified as 'underweight' were excluded from the statistical analysis. A final sample of 235 students divided into 11 classroom networks was achieved.

## Data collection
We collected data on sex, weight, height and social network contacts. We also asked about the physical exercise they carried out, including type and duration. Most of the data were collected via a questionnaire in print format administered between March and December 2015. Weight and height data were collected barefoot and minimal clothing. All data were collected by nursing staff trained for this purpose (members of the SALBIS (Salud, Bienestar, Ingeniería y Sostenibilidad Sociosanitaria) research group and the Castile and León regional health service, SACYL, Sanidad de Castilla y León). So as not to disrupt teaching, the questionnaire was administered during tutorial classes, and weight and height were measured during the physical education class. This latter procedure was carried out in a closed changing room using a portable Seca 700 stadiometer provided by the University of León Department of Nursing and Physiotherapy and electronic Fagor Slim scales calibrated to zero for each measurement.

To obtain data about social network contacts, classrooms were considered to represent entire networks and relationships within these networks were used in order to analyse the centrality of the individual within the network to which he or she belongs. Each questionnaire contained a closed list with the names and surnames of other classmates participating in the study, previously provided by the school management. Although defining a 'peer group' can be complex and impact on estimates of the effect of certain outcomes,[47] Gest was able to verify the validity of the use of classrooms as a reliable group measure in relation to children nominating each other.[48] Participants were given the following instruction: 'Using the list below, indicate how much time you spend with your classmates'. Thus, we used the approach employed in previous research and asked participants to rate contact intensity using a 5-point Likert scale where 1='we never spend time together' and 5='we are always together'. Other researchers have used the Likert scale to reflect the intensity of contact.[49–52] On this issue, non-reciprocal friendships have indeed been valued, as it has been shown that children have unilateral friendships and that it is therefore important to use friendship nomination instruments without restriction.[53] Furthermore, given that the directionality of friendships can influence the behaviour of others,[54] it has been found that in terms of health-related behaviour, long-term exposure to peers (such as in a classroom) can be correlated with an individual's behaviour.[55] All data were anonymised from the outset to ensure confidentiality, which is a standard procedure in social network studies.[56–60]

Physical activity levels and type of sport played were assessed using the Physical Activity Questionnaire for Adolescents (PAQ-A),[61] validated for the Spanish population (intraclass correlation coefficient=0.71, internal consistency α=0.74) between 12 and 17.[62] This belongs to a group of questionnaires known as the 'PAQ family'[62] and is considered a simple and reliable tool for assessing this type of activity in this population group. Through nine questions it asks about any physical activity carried out in the last 7 days. In the first eight questions, the participants are generally asked about the sport they did during the week and at the weekend, while in question number 9, which is not part of the test score, the young person is asked to specify whether he or she has been ill or something has prevented him or her from doing physical activity. Each question has five options for answers, which are equivalent to a score of 1–5. To obtain the equivalent data for this questionnaire, the score of each answer must be added up and divided by the total number of questions (8).

## Data analysis

Sex was considered a dichotomous variable. Each student's body mass index (BMI) was calculated from his or her height and weight. Furthermore, AnthroPlus software (WHO, Geneva, Switzerland), a tool provided by the WHO, was used to calculate percentiles according to exact age and sex and to classify each participant according to his or her weight status (underweight, healthy weight, overweight and obesity). We then generated a dichotomised variable using healthy weight as the reference category and combined overweight and obesity as the second category, which we termed overweight.

Social contact was analysed using UCINET V.6.365 (Analytic Technologies, Lexington, USA).[63] Data were used to generate an initial n×n matrix (single-mode or type I network), consisting of rows and columns of students belonging to each classroom network. Since we wished to study contact intensity, each frequency was assigned a score, which enabled us to create three different adjacency matrices (0, 1) from the initial matrix. They were based on three dichotomisation criteria: (A) a 'minimum contact' matrix, in which the original value of 1 ('we never spend time together') represents the absence of contact (0) and the values 2, 3, 4 and 5 ('we sometimes spend time together', 'we spend quite a lot of time together', 'we are almost always together' and 'we are always together') indicate the existence of the same (1); (B) an 'intermediate contact' matrix, in which the values 1 and 2 ('we never spend time together' and 'we sometimes spend time together') indicate the absence of contact (0) and the values 3, 4 and 5 ('we spend quite a lot of time together', 'we are almost always together' and 'we are always together') represent the existence of a tie (1); (C) a 'maximum contact' or 'friendship' matrix, where the values 1, 2 and 3 ('we never spend time together', 'we sometimes spend time together' and 'we spend quite a lot of time together') indicate a lack of contact (0), and 4 and 5 ('we are almost always together' and 'we are always together') represent the existence of a relationship (1).

For each contact intensity matrix, we analysed seven social network parameters: (1) outdegree (nominations emitted); (2) indegree (nominations received); (3) degree (number of ties that one actor has)[64]; (4 and 5) in/outcloseness (proximity, or number of steps that one actor must take to reach another)[65 66]; (6) betweenness (degree of connections that pass through an actor for one actor to reach another)[65 66] and (7) the eigenvector (a measure to identify the most central actors with the shortest distance to the rest of the nodes).[67] In combination, parameters (1)–(5) represent an individual's centrality.[68] As a result of this analysis, we obtained 21 normalised variables (values in which the ends are relativised carried out by the UCINET programme itself) organised by terciles (tercile 1 (T1), tercile 2 (T2) and tercile 3 (T3)).

Adolescents who answered question 9 of the PAQ-A Test in the affirmative (eg, that they had been unwell or something had prevented them from doing exercise in the last week) were excluded from the statistical analysis for all the variables related to physical activity but were not excluded from the nomination lists. The score for the level of physical activity measured by terciles. Students were grouped into three categories according to their level of activity, from lowest to highest. Thus, the first tercile (T1) indicated low activity and included students with scores between 1 and 2.33. The second tercile (T2) represented moderate activity and consisted of those who obtained scores between 2.34 and 2.80. The third tercile (T3) indicated high activity and was made up of students with scores between 2.81 and 5. Another variable created on student participation in sports was derived from question 1 on the PAQ-A questionnaire, and was identified as 'group sports'. This item contained a list of various sports, and students were asked to identify those they participated in and the frequency of their participation. Sports

Table 2  Descriptive data for BMI and percentiles according to the WHO, by sex.

|  | N | Mean | SD | Min | Max | P25 | Median | P75 |
|---|---|---|---|---|---|---|---|---|
| BMI | 235 | 22.1 | 2.9 | 17.4 | 35.4 | 20.0 | 21.5 | 23.8 |
| BMI males | 235 | 22.4 | 3.0 | 17.4 | 35.4 | 20.1 | 21.9 | 24.6 |
| BMI females | 235 | 21.7 | 2.7 | 17.8 | 32.4 | 19.9 | 21.4 | 23.4 |
| Percentile | 235 | 64.8 | 24.1 | 14.4 | 99.9 | 45.9 | 67.6 | 85.8 |
| Percentile males | 119 | 68.5 | 24.2 | 15.3 | 99.9 | 47.9 | 72.4 | 91.7 |
| Percentile females | 116 | 60.9 | 23.6 | 14.4 | 99.6 | 43.2 | 63.2 | 81.3 |

BMI, body mass index.

played in groups or teams (rugby, football, volleyball, hockey, basketball and handball) were used to distinguish students who played at least once or two times per week from those who never played. The latest was treated as a dichotomous variable.

In the statistical analysis, the relationship established between weight status and the study variables was determined by unconditional logistic regression. In each case, we calculated the OR with a 95% CI. The level of statistical significance (p value) of the values was established as p≤0.05. Statistical analyses were made using SPSS V.23 (IBM Corporation), and the centrality of the individual within the network were calculated using UCINET V.6.365 (Analytic Technologies, Lexington, USA).[69]

## RESULTS
### Descriptive
Of the total sample, 116 adolescents were female (49.4%) with a mean age of 15.4±0.8 years, and 119 were male (50.6%), with a mean age of 15.5±0.9 years. Ages ranged between 14 and 18.1. The mean BMI obtained was 22.1±2.9 kg/m$^2$, and the mean percentile value was 64.8±24.1. The results by sex are given in table 2.

According to the WHO reference standard,[70] 30.2% of the participants in our study were overweight (25.5% overweight and 4.7% obesity). By sex, the prevalence of being overweight and obese was 18.1% and 2.6%, among females, and 32.8% and 6.7%, among males (OR: 2.50; 95% CI: 1.40 to 4.47; p value: 0.002).

For physical activity, 39 individuals were excluded from the analyses because they were ill or indisposed, so the sample used to calculate the variables related to physical activity was 103 men (52.5%) and 93 women (47.5%). The mean score obtained from the PAQ-A (with a possible range of 1 to 5) was 2.6±0.6 (2.60±0.6 for males; 2.45±0.6 for females), corresponding to a moderate level of physical activity. Males obtained significantly higher scores in the PAQ-A Test (OR: 2.11; 95% CI: 1.04 to 4.25; p value: 0.036). Regarding group sports, 73% of respondents reported playing such sports, while 27% said that they did not participate in any of the group sports listed. Relative to females, males were more likely to participate in group sport (OR: 4.59: 95% CI: 2.28 to 9.22; p value: 0.000).

### Study of factors associated with overweight
We found no significant relationship between physical activity and weight status (table 3), but a more exhaustive analysis of the variables related to participation in sports revealed statistically significant results according to sex. We found a significant association in females between being overweight and reporting high levels of physical exercise (OR: 4.50; 95% CI: 1.21 to 16.74; p value: 0.025). As regards the variable 'group sports', we found no significant relationship between this and weight status, either in general or by sex.

### Study of the centrality of the individual within the network in terms of participation or not in group sports
We found a statistically significant association between participation in group sports and relationship parameters, both overall and by sex. In general, this association was significant at intermediate and maximum contact intensity levels. At both these levels of intensity, adolescent who participated in group sports were more likely to be classified as having low indegree centrality (popularity) (table 4). In the case of females, we obtained interesting data at all three contact intensity levels (minimum, intermediate and maximum). At the minimum contact intensity level, adolescent girls who participated in

Table 3  Study of the relationship between overweight and physical activity

|  | Normal weight | | Overweight | | OR | 95% CI | P value |
|---|---|---|---|---|---|---|---|
|  | N | % | N | % |  |  |  |
| Physical exercise | | | | | | | |
| Low activity | 47 | 72.3 | 18 | 27.7 | 1 | | |
| Moderate activity | 50 | 75.8 | 16 | 24.2 | 0.84 | 0.38 to 1.82 | 0.653 |
| High activity | 43 | 66.2 | 22 | 33.8 | 1.34 | 0.63 to 2.82 | 0.448 |

Table 4  OR between the variable 'group sports' and indegree (nominations received), at intermediate (companionship) and maximum (friendship) levels of intensity

| | Non-participant in group sports | | Participant in group sports | | OR | 95% CI | P value |
|---|---|---|---|---|---|---|---|
| | N | % | N | % | | | |
| Intermediate contact indegree | | | | | | | |
| T1 | 11 | 18.0 | 50 | 82.0 | 2.92 | 1.30 to 6.58 | **0.010** |
| T2 | 15 | 22.7 | 51 | 77.3 | 2.19 | 1.03 to 4.63 | **0.041** |
| T3 | 27 | 39.1 | 42 | 60.9 | 1 | | |
| Maximum contact indegree | | | | | | | |
| T1 | 15 | 22.1 | 53 | 77.9 | 2.17 | 1.01 to 4.68 | **0.047** |
| T2 | 14 | 21.5 | 51 | 78.5 | 2.24 | 1.03 to 4.90 | **0.042** |
| T3 | 24 | 38.1 | 39 | 61.9 | 1 | | |

Indegree: nominations received.
T1, tercile 1; T2, tercile 1; T3, tercile 1.

group sports were more likely to be classified as having low incloseness centrality (accessibility) and low eigenvector centrality (prestige) (table 5). At the intermediate contact intensity level, adolescent girls who participated in group sports were more likely to be classified as having low indegree centrality (popularity) and low eigenvector (prestige) (OR: 3.90; 95% CI: 1.26 to 12.08; p value: 0.019 and OR: 2.88; 95% CI: 1.02 to 8.13; p value: 0.046, respectively). At the maximum contact level, those adolescent girls who participated in group sports were more likely to be classified as having low indegree centrality (popularity) and low outcloseness centrality (communication skills)) (OR: 3.97; 95% CI: 1.30 to 12.13; p value: 0.016 and OR: 3.40; 95% CI: 1.22 to 9.49; p value: 0.019, respectively). In males, at the minimum intensity level, adolescent boys who participated in group sports were more likely to be classified as having high degree centrality (social activity) (OR: 5.12; 95% CI: 0.99 to 26.27; p value: 0.050). Figure 1 shows this effect. The bigger nodes represent a greater centrality according to the eigenvector, the squares are male while the round ones represent the female sex.

The red nodes indicate the performance of group sports while the white nodes indicate not carrying out this type of sport. It can be observed how the red square nodes (males doing group sports) are larger in comparison with the red round nodes (females doing group sports), thus indicating that they have a greater centrality.

However, we did not obtain a significant association between participation in group sports and any of the network parameters in the group of students classified as overweight.

## DISCUSSION

It has been widely reported in the literature that physical exercise has multiple health benefits for adolescents.[71] High intensity physical activity not only improves their biopsychosocial health but also influences their weight status.[72 73] Despite these findings, a literature review by Borraccino et al[74] revealed that adolescents from many countries did minimal exercise, and did not meet the recommendations given in various published guidelines.

Table 5  OR between female participation in group sports and incloseness (proximity) and the eigenvector (prestige) at minimum contact intensity level

| | Non-participant in group sports | | Participant in group sports | | OR | 95% CI | P value |
|---|---|---|---|---|---|---|---|
| | N | % | N | % | | | |
| Incloseness | | | | | | | |
| T1 | 10 | 32.3 | 21 | 67.7 | 2.90 | 1.03 to 8.20 | **0.044** |
| T2 | 11 | 35.5 | 20 | 64.5 | 2.52 | 0.90 to 7.01 | 0.077 |
| T3 | 18 | 58.1 | 13 | 41.9 | 1 | | |
| Eigenvector | | | | | | | |
| T1 | 11 | 33.3 | 22 | 66.7 | 3.27 | 1.15 to 9.28 | **0.026** |
| T2 | 10 | 32.3 | 21 | 67.7 | 3.44 | 1.19 to 9.95 | **0.023** |
| T3 | 18 | 62.1 | 11 | 37.9 | 1 | | |

In/Outcloseness: degree of proximity; Eigenvector: degree of prestige/influence.
T1, tercile 1; T2, tercile 1; T3, tercile 1.

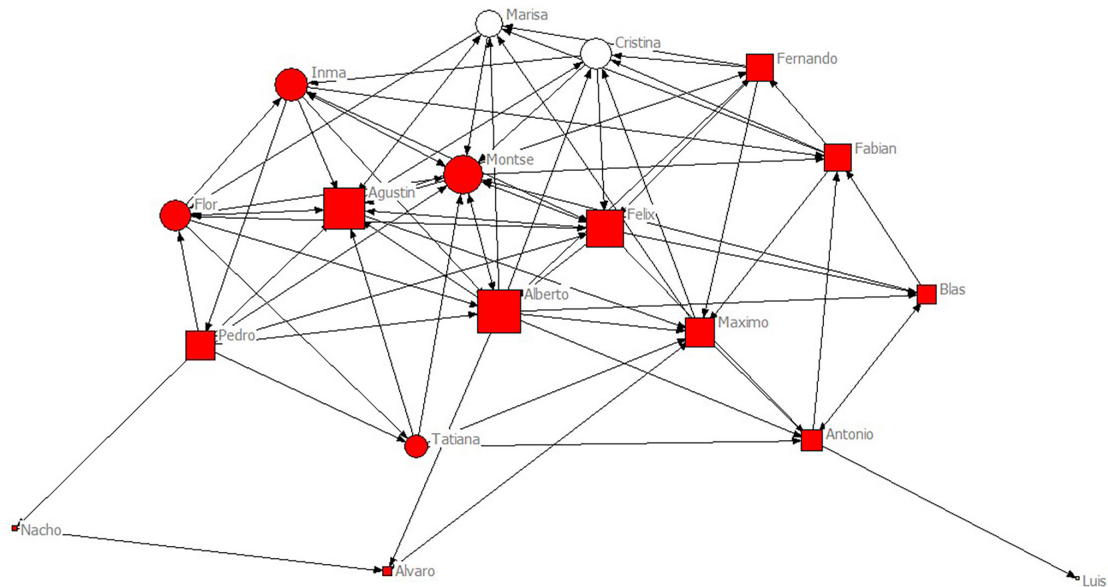

**Figure 1** The squares represent the men and the circles represent the women. The colour red indicates group sport, while the colour white indicates no group sport. Node size indicates prestige by measuring the eigenvector. The contact level represented is the intermediate contact level. The names of each individual are fictitious. The graphics were produced using UCINET software.[69]

Furthermore, adolescence is also the stage when young people are at the highest risk of giving up sport, since in general physical activity levels decrease with age within the adolescent age group,[75] or also in the adolescent—young adult age group.[76] This situation is aggravated if there is also an increase in inactivity with age overall.[77]

In this study, the mean level of physical activity in the current sample corresponded to moderate activity levels, although males were more likely to report high levels of physical activity and group sport involvement than females. We did not find a significant relationship between weight status and physical exercise in the total sample, but did find one by sex: in females, higher amounts of physical exercise were related to being overweight. There are three possible explanations for this finding. First of all physical activity levels are not the exclusive cause of overweight in adolescents. Another is associated with a finding reported by McMurray et al,[78] whereby overweight adolescents tended to overestimate their physical activity levels. Thus, when their physical activity was measured objectively (via an accelerometer) and the results compared with their questionnaire responses, it was found that overweight adolescents tended to report higher activity than was the case. As in the case of diet, a third explanation is reverse causality, an effect in cross-sectional studies,[79] whereby overweight females may do more exercise to achieve their ideal figure to achieve improved self-esteem related to the body image that they usually perceive.[80] However, the fact that BMI was used as an indicator of weight status may justify this result, as, although it is a widely accepted indicator by other researchers. BMI does not allow evaluating differences in lean body mass and fat mass between athletes and non-athletes.[81]

The results obtained from analysing group sports show that 73% of the sample reported doing some group sport, such as rugby, football, volleyball, hockey, basketball or handball. By sex, males were significantly related to participation in group sports. In this sense, Klomsten et al[82] also find in their study a greater predilection of men for sports such as football or ice hockey or even individual sports with a higher burden of violence such as boxing. On the other hand, adolescent girls choose activities such as dance or aerobics. These activities, although not necessarily less active than others, seem to be marked by gender stereotypes. There is also the fact that adolescent girls who carry out these types of sports, which are not affiliated to a sports federation, are not considered by their environment to be active in sport. However, Vilhjamsson et al[83] support the idea of a greater number of boys enrolled in organised sports in the same vein as the results obtained in this work, that is, boys outnumber girls in competitive, team and high-intensity physical activity, while girls do so in individual, non-competitive and medium-low intensity activities, such as walking, aerobic gymnastics, skating.

However, as with the amount of physical exercise, we did not find an association between excess weight and participation in group sports, either in general or by sex. These findings are contrary to those obtained by other researchers. For example, Weintraub et al[84] found that following an intervention consisting of football activities as an after-school sport, this type of group sport presented good acceptance and good adherence, and managed to lower the BMI of children within a period of 3–6 months. Other researchers[85 86] found that group sports were inversely associated with BMI.

To explore the social aspect of physical activity, we also analysed the position of the adolescent within the network of adolescents who participated in group sports. Our results show that participation in group sports was related to low values for nominations received at intermediate and maximum contact intensity levels, thus showing a lower popularity in individuals who have more contact, compared with individuals who do not participate in group sport. In contrast, Devecioglu et al[87] found that university students who participated in group sports presented higher levels of social relations, with no differences by sex. Nevertheless, these results invite us to reflect on the use of female athletes as influencers, as other researchers have stated,[44] since the results obtained in our work would make it difficult to establish a mechanism of contagion between peers. The authors believe that our results are the result of a deeper study into contact levels and that this should be taken into account in future strategies related to influencing agents in physical activity.

The data obtained by sex are even more striking since they indicate that for adolescents, the social impact/benefit of participating in group sports depends on their sex. At minimum contact level, female adolescents who participated in this type of sport were distant from their other peers and had a low level of prestige. At an intermediate contact level, they obtained low values for nominations received and had little prestige, while at maximum contact level (friendship), they again obtained low values for nominations received as well as low values for the possibility of reaching their peers (proximity). Therefore, they were being isolated from their peers. However, the opposite was the case of the males. At a minimum contact level, this participation was related to having a higher number of contacts or greater social activity. Similarly, de la Haye et al[42] found that males who participated in organised physical activities (supervised by a trainer/adult) were more popular in their classroom network. Contrary to our results, in the study conducted by Diaconu-Gherasim and Duca,[88] they found that compared with males, girls who played group sports reported greater social benefits in the relationship established with their best friends and were generally more socially competent. Nevertheless, as a possible explanation for the results we have obtained, we consider the mocking and aspects related to body image among other types of barriers,[89 90] among others as a possible justification for the social effect found.

The way in which relationships are established according to the type of physical activity undertaken has received little research attention in SNA. Nonetheless, friendship has been analysed in terms of its influence on participation in physical exercise[25 30 31] and levels of physical activity.[31] Also, the mechanism whereby friendship is formed around sport has been studied to determine the extent to which physical activity shapes or is shaped by adolescent friendships.[36] In this respect, our sample indicates that the process may be different depending on sex, whereby sport facilitates social relations for males, but some sports present a transparent barrier to the same for females.

## Limitations

As important limitations found we underline the importance of the sample size. This could be a reason for the inconsistent findings of our study compared with the literature reviewed. We also consider it a limitation to consider a social network as that generated within the school environment, without taking into account family factors and friendship connections outside school, which, we are aware, are also important for the adolescent. Similarly, it may be biased not to have included in the closed list provided to adolescents to nominate their peers, students with whom they share a classroom but who have not completed the informed consent form. Future studies should also consider the selection of the course as a network to be studied and not to establish networks by classrooms. On the other hand, the cross-sectional design of the study prevents analysis with respect to the mechanisms by which the peers set out in the introduction section (selection vs infection) are grouped together. Finally, the limitation that occurs with self-reported data is recognised.

## CONCLUSIONS

Our findings, although tenuous, underline the importance of two fundamental aspects to be taken into account in the design of future strategies to promote physical activity in adolescence: gender and the consequences on study of centrality within their social network according to the intensity of contact they have with their peers. SNA offers multiple possibilities to learn what happens and is transferred among adolescents. Networks, thus, represent a handy tool for instilling healthy behaviour, based on information provided by the study of the structure of the network and not only the characteristics of the individual. This study paves the way for new lines of research that support, enrich and further this relational perspective, which could help promote healthy habits in adolescents.

**Twitter** José Alberto Benítez-Andrades @jabenitez88

**Contributors** Conceptualisation, NA and JAB-A; methodology, PM-S and MDCS; software, JAB-A; validation, NA, MDCS and PM-S; formal analysis, JAB-A and NA; investigation, NA; resources, MDCS ; data curation, JAB-A and NA; writing–original draft preparation, NA and JAB-A; writing–review and editing, NA; visualisation, JAB-A; supervision, PM-S and MDCS; project administration, NA; funding acquisition, PM-S. All authors have read and agree to the published version of the manuscript.

**Funding** The authors have not declared a specific grant for this research from any funding agency in the public, commercial or not-for-profit sectors.

**Competing interests** None declared.

**Patient and public involvement** Patients and/or the public were involved in the design, or conduct, or reporting, or dissemination plans of this research. Refer to The study section for further details.

**Patient consent for publication** Parental/guardian consent obtained.

**Ethics approval** Permission for data collection was sought from the Castile and León Education Department and the Spanish Data Protection Agency (CV: BOCYL-D-12032015-6), and interviews were conducted with heads of school and class tutors to obtain their consent and collaboration. All students were given an informed consent form in a sealed envelope addressed to their parents. Lastly, parents were offered the possibility of retracting consent once they had signed the form, without needing to provide a reason, and an email contact address was given should they require any further information. Participation was voluntary, and subject availability was respected at all times.

**Provenance and peer review**  Not commissioned; externally peer reviewed.

**Data availability statement**  No data are available. No additional data available.

**ORCID iD**
José Alberto Benítez-Andrades http://orcid.org/0000-0002-4450-349X